\RequirePackage{lineno}

\documentclass[%
  10pt,
  aps,
  prl,
  twocolumn,
  superscriptaddress,
  showpacs,
  altaffilletter,
  amsmath,
  amssymb
]{revtex4-2}

\usepackage{graphicx}
\usepackage{bm}
\usepackage{multirow} 
\usepackage{siunitx} 

\usepackage{comment} 

\usepackage{xspace}
\usepackage{upgreek}
\usepackage[nodayofweek,level]{datetime}
\newdate{aug10}{10}{08}{2023}

\begin{document}
\setlength{\abovedisplayskip}{5pt}   
\setlength{\belowdisplayskip}{5pt}

\newcommand{\fix}[1]{\textcolor{red}{#1}\xspace}
\newcommand{\needref}{[\textcolor{red}{ref}]\xspace}
\newcommand{\amu}{\ensuremath{a^{}_{\mu}}\xspace}
\renewcommand{\ae}{\ensuremath{a^{}_{e}}\xspace}
\newcommand{\gm}{\ensuremath{g\!-\!2}\xspace}
\newcommand{\gmtwo}{\ensuremath{g\!-\!2}\xspace}
\newcommand{\wa}{\ensuremath{\omega_{a}}\xspace}
\newcommand{\wam}{\ensuremath{\omega_{a}^{m}}\xspace}
\renewcommand{\wp}{\ensuremath{\omega_{p}}\xspace}
\newcommand{\oa}{\ensuremath{\omega^{}_a}\xspace}
\newcommand{\ocycl}{\ensuremath{\omega^{}_c}\xspace}
\newcommand{\oS}{\ensuremath{\omega^{}_S}\xspace}
\newcommand{\op}{\ensuremath{\omega^{}_p}\xspace}
\newcommand{\opmeas}{\omega_{p}^{\pp}\xspace}
\newcommand{\opprime}{\ensuremath{\omega'^{}_p}\xspace}
\newcommand{\opprimeatTexp}{\ensuremath{\omega'^{}_p(\Tr)}\xspace}
\newcommand{\opprimetilde}{\ensuremath{\tilde{\omega}'^{}_p}\xspace}
\newcommand{\opprimetildeofT}{\ensuremath{\tilde{\omega}'^{}_p(T)}\xspace}
\newcommand{\opprimetildeatTexp}{\ensuremath{\tilde{\omega}'^{}_p(\Tr)}\xspace}
\newcommand{\opprimetildebold}{\ensuremath{\bm{\tilde{\omega}'^{}_p}}\xspace}
\newcommand{\vecr}{\ensuremath{\vec{r}}\xspace}
\newcommand{\weightedfield}{\ensuremath{f_{\rm calib}\,{\cdot}\,\langle \omega'^{}_p \times M \rangle}\xspace}
\newcommand{\weightedfieldbold}{\ensuremath{\bm{f_{\rm calib}\,{\cdot}\,\langle \omega'^{}_p \times M \rangle}}\xspace}
\newcommand{\optilde}{\ensuremath{\tilde{\omega}^{}_p}\xspace}
\newcommand{\omegap}{\op}
\newcommand{\omegapfree}{\ensuremath{\omega_p^\mathrm{free}}\xspace}
\newcommand{\fid}{FID\xspace}
\newcommand{\BNLexperiment}{\ac{BNL} E821 experiment\xspace}
\newcommand{\FNALexperiment}{\ac{FNAL} Muon \gm Experiment\xspace}
\newcommand{\RunOne}{Run-1\xspace}
\newcommand{\RunFourFiveSix}{Run-456\xspace}
\newcommand{\RunTwo}{Run-2\xspace}
\newcommand{\RunThree}{Run-3\xspace}
\newcommand{\RunTwoThree}{Run-2/3\xspace}
\newcommand{\RunOneTwoThree}{Run-1/2/3\xspace}
\newcommand{\RunThreeA}{Run-3a\xspace}
\newcommand{\RunThreeB}{Run-3b\xspace}
\newcommand{\RunFour}{Run-4\xspace}
\newcommand{\RunOneA}{Run-1a\xspace}
\newcommand{\RunOneB}{Run-1b\xspace}
\newcommand{\RunOneC}{Run-1c\xspace}
\newcommand{\RunOneD}{Run-1d\xspace}
\newcommand{\hethree}{\ensuremath{^3\text{He}}}
\newcommand{\water}{\ensuremath{\text{H_2O}}}
\newcommand{\pp}{\mathrm{cp}}

\newcommand{\opprimeq}{\ensuremath{\omega'^{}_{p,q}}\xspace}
\newcommand{\opprimej}{\ensuremath{\omega'^{}_{p,j}}\xspace}
\newcommand{\opprimeqj}{\ensuremath{\omega'^{}_{p,q,j}}\xspace}

\newcommand{\opprimetildeq}{\ensuremath{\tilde\omega'^{}_{p,q}}\xspace}
\newcommand{\opprimetildeqj}{\ensuremath{\tilde\omega'^{}_{p,q,j}}\xspace}
\newcommand{\Tr}{\ensuremath{T^{}_{r}\xspace}}

\newcommand{\sigmamuq}{\sigma^\mu_{q}}
\newcommand{\sigmamuqj}{\sigma^\mu_{q,j}}

\newcommand{\mi}{\ensuremath{m^{}_{i}}\xspace}
\newcommand{\mone}{\ensuremath{m^{}_{1}}\xspace}
\newcommand{\mtwo}{\ensuremath{m^{}_{2}}\xspace}
\newcommand{\mthree}{\ensuremath{m^{}_{3}}\xspace}
\newcommand{\mfour}{\ensuremath{m^{}_{4}}\xspace}
\newcommand{\mfive}{\ensuremath{m^{}_{5}}\xspace}
\newcommand{\msix}{\ensuremath{m^{}_{6}}\xspace}
\newcommand{\mseven}{\ensuremath{m^{}_{7}}\xspace}
\newcommand{\meight}{\ensuremath{m^{}_{8}}\xspace}
\newcommand{\mnine}{\ensuremath{m^{}_{9}}\xspace}
\newcommand{\mten}{\ensuremath{m^{}_{10}}\xspace}
\newcommand{\meleven}{\ensuremath{m^{}_{11}}\xspace}
\newcommand{\mtwelve}{\ensuremath{m^{}_{12}}\xspace}
\newcommand{\mthirteen}{\ensuremath{m^{}_{13}}\xspace}

\newcommand{\authornote}[1]{{%
  \let\thempfn\relax
  \footnotetext[0]{$\diamond${ }#1}
}}

\newcommand{\Rmu}{\ensuremath{{\mathcal R}_\mu}\xspace}
\newcommand{\Rmuprime}{\ensuremath{{\mathcal R}'^{}_\mu}\xspace}

\newcommand{\RmuprimeOne}{\ensuremath{{\mathcal R}'^{}_{\mu}(\text{Run-1})}\xspace}
\newcommand{\RmuprimeTwoThree}{\ensuremath{{\mathcal R}'^{}_{\mu}(\text{Run-2/3})}\xspace}
\newcommand{\RmuprimeOneTwoThree}{\ensuremath{{\mathcal R}'^{}_{\mu}(\text{Run-1/2/3})}\xspace}

\newcommand\Tstrut{\rule{0pt}{2.6ex}}         
\newcommand\Bstrut{\rule[-0.9ex]{0pt}{0pt}}   

\newcommand{\rmagic}{\ensuremath{r_{0}}\xspace}
\newcommand{\cmagic}{\ensuremath{c_{0}}\xspace}
\newcommand{\pmagic}{\ensuremath{p_{0}}\xspace}
\newcommand{\gammamagic}{\ensuremath{\gamma_{0}}\xspace}
\newcommand{\betamagic}{\ensuremath{\beta_{0}}\xspace}
\newcommand{\dipbfield}{\ensuremath{B_{0}}\xspace}
\newcommand{\runone}{Run-1\xspace}
\newcommand{\runtwo}{Run-2\xspace}
\newcommand{\runthree}{Run-3\xspace}
\newcommand{\runfour}{Run-4\xspace}
\newcommand\runonea{Run-1a\xspace}
\newcommand\runoneb{Run-1b\xspace}
\newcommand\runonec{Run-1c\xspace}
\newcommand\runoned{Run-1d\xspace}

\newcommand{\precession}{precession-run1}
\newcommand{\field}{field-run1}
\newcommand{\BD}{BD-run1}
 
\title{Measurement of the Positive Muon Anomalous Magnetic Moment to 0.20\,ppm}

%
\affiliation{Argonne National Laboratory, Lemont, Illinois, USA}
\affiliation{Boston University, Boston, Massachusetts, USA}
\affiliation{Brookhaven National Laboratory, Upton, New York, USA}
\affiliation{Budker Institute of Nuclear Physics, Novosibirsk, Russia}
\affiliation{Center for Axion and Precision Physics (CAPP) / Institute for Basic Science (IBS), Daejeon, Republic of Korea}
\affiliation{Cornell University, Ithaca, New York, USA}
\affiliation{Fermi National Accelerator Laboratory, Batavia, Illinois, USA}
\affiliation{INFN, Laboratori Nazionali di Frascati, Frascati, Italy}
\affiliation{INFN, Sezione di Napoli, Naples, Italy}
\affiliation{INFN, Sezione di Pisa, Pisa, Italy}
\affiliation{INFN, Sezione di Roma Tor Vergata, Rome, Italy}
\affiliation{INFN, Sezione di Trieste, Trieste, Italy}
\affiliation{Department of Physics and Astronomy, James Madison University, Harrisonburg, Virginia, USA}
\affiliation{Institute of Physics and Cluster of Excellence PRISMA+, Johannes Gutenberg University Mainz, Mainz, Germany}
\affiliation{Joint Institute for Nuclear Research, Dubna, Russia}
\affiliation{Department of Physics, Korea Advanced Institute of Science and Technology (KAIST), Daejeon, Republic of Korea}
\affiliation{Lancaster University, Lancaster, United Kingdom}
\affiliation{Michigan State University, East Lansing, Michigan, USA}
\affiliation{North Central College, Naperville, Illinois, USA}
\affiliation{Northern Illinois University, DeKalb, Illinois, USA}
\affiliation{Regis University, Denver, Colorado, USA}
\affiliation{School of Physics and Astronomy, Shanghai Jiao Tong University, Shanghai, China}
\affiliation{Tsung-Dao Lee Institute, Shanghai Jiao Tong University, Shanghai, China}
\affiliation{Institut f\"ur Kern- und Teilchenphysik, Technische Universit\"at Dresden, Dresden, Germany}
\affiliation{Universit\`a del Molise, Campobasso, Italy}
\affiliation{Universit\`a di Udine, Udine, Italy}
\affiliation{Department of Physics and Astronomy, University College London, London, United Kingdom}
\affiliation{University of Illinois at Urbana-Champaign, Urbana, Illinois, USA}
\affiliation{University of Kentucky, Lexington, Kentucky, USA}
\affiliation{University of Liverpool, Liverpool, United Kingdom}
\affiliation{Department of Physics and Astronomy, University of Manchester, Manchester, United Kingdom}
\affiliation{Department of Physics, University of Massachusetts, Amherst, Massachusetts, USA}
\affiliation{University of Michigan, Ann Arbor, Michigan, USA}
\affiliation{University of Mississippi, University, Mississippi, USA}
\affiliation{University of Virginia, Charlottesville, Virginia, USA}
\affiliation{University of Washington, Seattle, Washington, USA}
\author{D.~P.~Aguillard}  \affiliation{University of Michigan, Ann Arbor, Michigan, USA}
\author{T.~Albahri}  \affiliation{University of Liverpool, Liverpool, United Kingdom}
\author{D.~Allspach}  \affiliation{Fermi National Accelerator Laboratory, Batavia, Illinois, USA}
\author{A.~Anisenkov} \altaffiliation[Also at ]{Novosibirsk State University.}  \affiliation{Budker Institute of Nuclear Physics, Novosibirsk, Russia}
\author{K.~Badgley}  \affiliation{Fermi National Accelerator Laboratory, Batavia, Illinois, USA}
\author{S.~Bae{\ss}ler} \altaffiliation[Also at ]{Oak Ridge National Laboratory.}  \affiliation{University of Virginia, Charlottesville, Virginia, USA}
\author{I.~Bailey} \altaffiliation[Also at ]{The Cockcroft Institute of Accelerator Science and Technology, Daresbury, United Kingdom.}  \affiliation{Lancaster University, Lancaster, United Kingdom}
\author{L.~Bailey}  \affiliation{Department of Physics and Astronomy, University College London, London, United Kingdom}
\author{V.~A.~Baranov} \thanks{Deceased.} \affiliation{Joint Institute for Nuclear Research, Dubna, Russia}
\author{E.~Barlas-Yucel}  \affiliation{University of Illinois at Urbana-Champaign, Urbana, Illinois, USA}
% ...
\collaboration{The Muon \gmtwo Collaboration} \noaffiliation
\vskip 0.25cm

 \authornote{Deceased}
\date{\displaydate{aug10}}

\begin{abstract}

We present a new measurement of the positive muon magnetic anomaly, $\amu \equiv (g_\mu - 2)/2$, from the Fermilab Muon \gmtwo Experiment using data collected in 2019 and 2020.
We have analyzed more than 4 times the number of positrons from muon decay than in our previous result from 2018 data.
The systematic error is reduced by more than a factor of 2 due to better running conditions, a more stable beam, and improved knowledge of the magnetic field weighted by the muon distribution, \opprimetilde, and of the anomalous precession frequency corrected for beam dynamics effects, \wa.
From the ratio $\wa / \opprimetilde$, together with precisely determined external parameters, we determine $\amu = 116\,592\,057(25) \times 10^{-11}$~(0.21~ppm). Combining this result with our previous result from the 2018 data, we obtain $\amu\text{(FNAL)} = 116\,592\,055(24) \times 10^{-11}$~(0.20~ppm). The new experimental world average is $a_\mu (\text{exp}) = 116\,592\,059(22)\times 10^{-11}$~(0.19~ppm), which represents a factor of 2 improvement in precision.
\end{abstract}

\maketitle 


\textit{Introduction.}---A precise experimental measurement of the muon magnetic anomaly \amu provides a stringent test of the Standard Model (SM) as it can be theoretically predicted with high precision.
Any deviation between experiment and theory may be a sign of physics beyond the SM. 
We report a new measurement of \amu using data collected in 2019 (\runtwo) and 2020 (\runthree) by the Muon \gmtwo Experiment at Fermilab.
The data constitute a fourfold increase in detected positrons compared to our previous measurement (\RunOne) ~\cite{Run1PRL,Run1PRDomegaa,Run1PRAField,Run1PRABBeamDyn}.
Analysis and run condition improvements also lead to more than a factor of 2 reduction in the systematic uncertainties, surpassing the experiment's design goal~\cite{TDR}.

Our Run-1 publications describe the principle of the experiment, previous results, and experimental details~\cite{Run1PRL,Run1PRDomegaa,Run1PRAField,Run1PRABBeamDyn}.
The experiment uses $3.1$~GeV/c polarized muons produced at the Fermilab Muon Campus~\cite{muoncampus}.
Muons are injected into a $7.112$-m radius storage ring that was moved, and significantly upgraded, from the BNL experiment~\cite{Danby:2001eh, BNLFinalReport}.
Two key components of the storage ring are kicker magnets that direct the injected muons onto the central orbit of the storage ring~\cite{kickerpaper} and electrostatic quadrupoles (ESQs) that provide vertical focusing of the stored beam~\cite{e821quadpaper}.  
The anomalous spin precession frequency \wa---the difference between the muon spin precession frequency and the cyclotron frequency---is measured by recording the time dependence of the number of high-energy positrons detected in a series of calorimeters located on the inner radius of the storage ring~\cite{calopaper}.
The magnetic field is mapped every few days using a trolley instrumented with nuclear magnetic resonance (NMR) probes housing petroleum jelly~\cite{Corrodi_2020}.
These probes are calibrated using a retractable water-based cylindrical probe~\cite{plungingprobepaper}.
This enables the expression of the magnetic field in terms of the precession frequency of shielded protons in a spherical sample $\opprime$, for which the relation between precession frequency and magnetic field is precisely known.
After weighting for the muon spatial distribution, the precession frequency is denoted $\opprimetilde$.
Changes in the field between trolley measurements are tracked using NMR probes embedded in the vacuum chamber walls above and below the muon storage volume ~\cite{Run1PRAField}.
Dedicated instrumentation is used to measure transient magnetic fields caused by the pulsing of the kickers and ESQs.
The spatial distribution of the muon beam within the storage ring as a function of time since injection is inferred from positron trajectories recorded using two tracking detectors~\cite{trackerpaper}.

We incorporated major instrumental improvements with respect to \RunOne.
Resistors in the high-voltage feedthroughs for the ESQ system that were damaged in \RunOne were replaced before \RunTwo.
This upgrade greatly improved transverse beam stability.
Thermal insulation was added to the storage ring magnet before \RunTwo to remove diurnal temperature variations.
Increased cooling power and improved air circulation in the experimental hall installed before \RunThree reduced seasonal temperature variations.
The magnitude and reliability of the kicker field were improved between \RunOne and \RunTwo, and again within \RunThree.
Because of these improvements, the data are analyzed in three sets---\RunTwo, \RunThreeA, and \RunThreeB. A full description of the hardware upgrades, operating conditions and analysis details will be provided in an in-depth paper currently in preparation.

The data are blinded by hiding the true value of the calorimeter digitization clock frequency.
This blinding factor is different for \RunTwo and \RunThree. 

We obtain the muon magnetic anomaly from~\footnote{We use the shielded proton-to-electron magnetic moment ratio~\cite{Phillips:1977} and the electron $g$-factor~\cite{electronge, pdgelectronge} measurement.
The CODATA-2018 result is used for the muon-to-electron mass ratio~\cite{CODATA:2018}, which is determined from bound-state QED theory and measurements described in Ref.~\cite{Liu:1999iz}.
The QED factor $\mu_e(H)/\mu_e$ is computed by theory with negligible uncertainty~\cite{CODATA:2018}.}
\begin{linenomath*}
\begin{equation}
\begin{aligned}
a_{\mu} = \frac{\wa}{\opprimetildeatTexp} \frac{\mu'_p(T_{r})}{\mu_e(H)} \frac{{\mu_e(H)}}{\mu_e} \frac{m_{\mu}}{m_e} \frac{g_e}{2},
\label{eq:amueq}
\end{aligned}
\end{equation}
\end{linenomath*}

where this experiment measures two frequencies to form the ratio $\Rmu^{'} =\wa/\opprimetildeatTexp$,
where $T_r = 34.7$~$^{\circ}$C is the temperature at which the shielded proton-to-electron magnetic moment is measured \cite{Phillips:1977}.
The ratio of the measured frequencies must be corrected for a number of effects, which shift the value of \Rmuprime by $+$622~ppb in total.
We write the ratio in terms of measured quantities and corrections as
\begin{linenomath*}
\begin{equation}
\Rmu^{'}  \approx \frac{f_{\rm clock}{\cdot}\,\wam  \left (1 + C_e + C_p + C_{pa} + C_{dd} + C_{ml}   \right )}{f_{\rm calib}\,{\cdot}\,\langle \opprime(\vecr)\times M(\vecr)\rangle (1+B_{q}+B_{k}) }.
\label{eq:Rcomponents}
\end{equation}
\end{linenomath*}
The numerator consists of the clock-blinding factor $f_{\text{clock}}$, the measured precession frequency \wam, and five corrections $C_i$ associated with the spatial and temporal motion of the beam.
In the denominator, we separate \opprimetildeatTexp into the absolute NMR calibration procedure (indicated by $f_{\rm calib}$) and the magnetic field maps, which are weighted by the muon spatial distribution and positron count [$\langle \opprime(\vecr)\times M(\vecr)\rangle$, where the average is over all points \vecr within the storage region].
We apply corrections $B_i$ to the magnetic field to account for two fast magnetic transient fields that are synchronized to the muon storage period.
The uncertainties and correction values for the elements of Eq.~\ref{eq:Rcomponents} are shown in Table~\ref{tb:systematics}.

\begin{table}
\begin{ruledtabular}
\begin{tabular}{lcc}
\multirow{2}{*}{Quantity} & Correction & Uncertainty\\
                          & (ppb)            & (ppb)\Bstrut\\
\hline
\wam\ (statistical) & - & 201\Tstrut\\
\wam\ (systematic)  & - & 25\Bstrut\\
\hline
$C_e$     & 451 & 32\Tstrut\\
$C_p$     & 170 & 10\\
$C_{pa}$  & $-27$ & 13\\
$C_{dd}$  & $-15$ & 17\\
$C_{ml}$  &   0 &  3\Bstrut\\
\hline
$f_{\text{calib}}\,{\cdot}\,\langle \opprime(\vecr)\times M(\vecr)\rangle$ & - & 46\Tstrut\\
$B_{k}$   & $-21$ & 13 \\
$B_{q}$   & $-21$ & 20\Bstrut\\
\hline
$\mu'_p(34.7^\circ)/\mu_e$  & - & 11\Tstrut\\
$m_\mu/m_e$  & - & 22\\
$g_e/2$ & - & 0\Bstrut\\
\hline
Total systematic for \Rmuprime & - & 70\Tstrut\\
Total external parameters & - & 25 \\
Total for \amu & 622 & 215\Bstrut\\
\end{tabular}
\end{ruledtabular}
\caption{Values and uncertainties of the \Rmuprime terms in Eq.~\ref{eq:Rcomponents}, and uncertainties due to the external parameters in Eq.~\ref{eq:amueq} for \amu. Positive $C_i$ increases \amu; positive $B_i$ decreases \amu (see Eq.~\ref{eq:Rcomponents}). The \wam uncertainties are decomposed into statistical and systematic contributions. All values are computed with full precision and then rounded to the reported digits.}
\label{tb:systematics}
\end{table}


\textit{Anomalous precession frequency }\wam.---The time dependence of the number of positrons from muon decays recorded by calorimeters in a storage period is given by
\begin{linenomath*}
\begin{multline}
N(t)=N_0\eta_{N}(t) e^{-t / \gamma\tau_\mu} \\
\times\left\{1+A \eta_{A}(t)\cos\left[\wam t + \varphi_0 + \eta_{\phi}(t)\right]\right\},
	\label{eq:wiggle_func}
\end{multline}
\end{linenomath*}
where $N_0$ is the normalization, $\gamma\tau_\mu$ is the time-dilated muon lifetime ($\approx64.4\,\upmu$s), $A$ is the average weak-decay asymmetry, and $\varphi_0$ is the average phase difference between the muon momentum and spin directions at the time of muon injection. The normalization, asymmetry, and phase have time-dependent correction factors, $\eta_N$, $\eta_A$, and $\eta_\phi$, that account for horizontal $(x)$ and vertical $(y)$ beam oscillations, including $x$-$y$ coupling.

Nearly all parameters in Eq.~\ref{eq:wiggle_func} have some energy dependence, but it is particularly strong for $N_0$ and $A$.
We choose to combine the data in the statistically optimal way of weighting each positron by its energy-dependent asymmetry~\cite{Bennett:2007zzb}.

\begin{figure}[ht]
\centering
\includegraphics[width=\columnwidth]{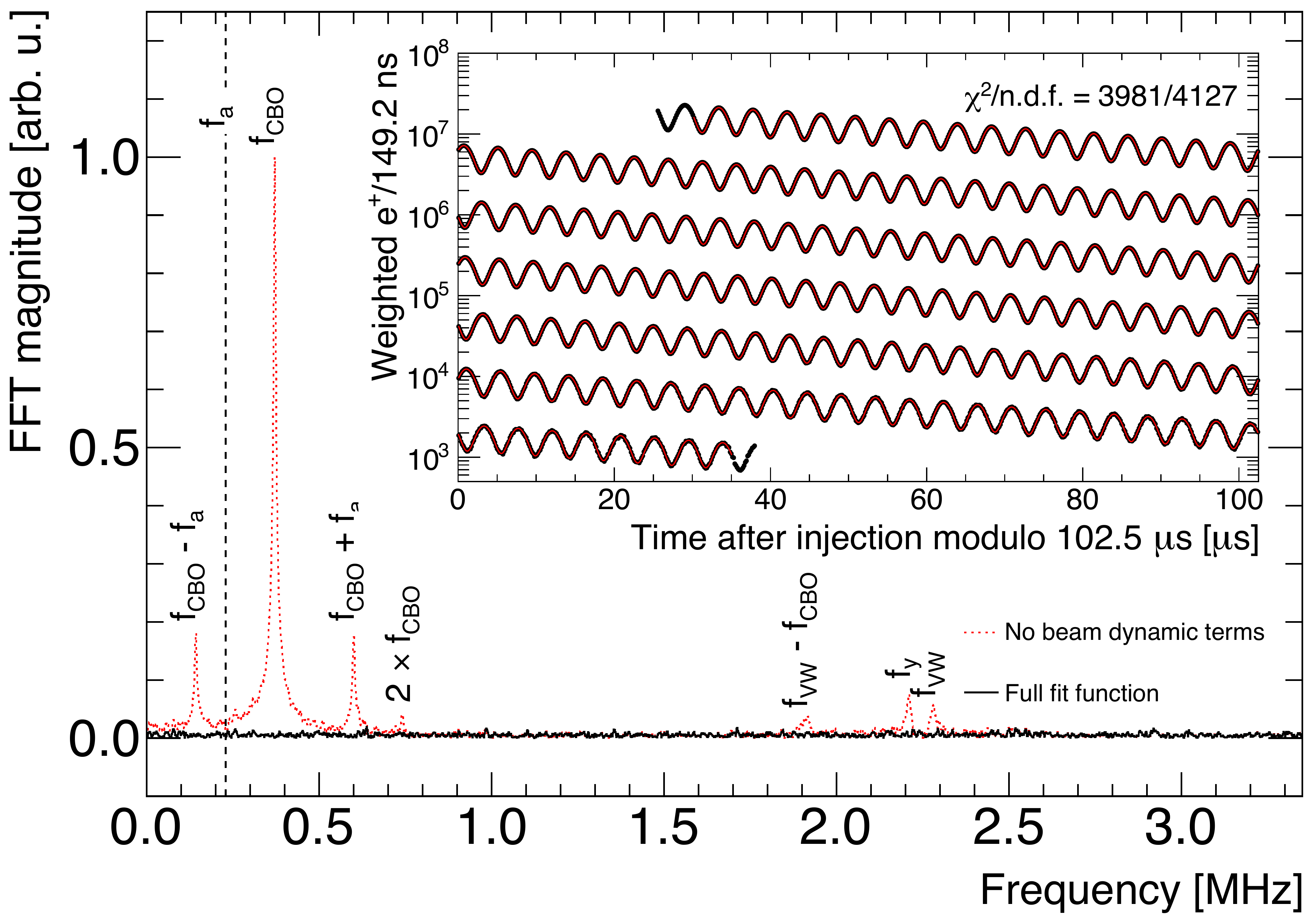}
\caption{Fourier transform of the residuals from a fit following Eq.~\ref{eq:wiggle_func} excluding $\eta_N$, $\eta_A$, and $\eta_\phi$ (red dashed line), and from the full fit (black line).  The peaks correspond to the missing betatron frequencies and muon losses.  Data are from the \RunThreeA data set.
Inset: corresponding asymmetry-weighted $e^{+}$ time spectrum (black line) with the full fit function (red line) overlaid.}
\label{fig:precession}
\end{figure}

Seven different analysis groups perform independent extractions of \wam by a $\chi^{2}$ minimization.
Each analysis team adds an independent blind offset to their result in addition to the aforementioned clock blinding.
Two groups perform a new asymmetry-weighted ratio method by subdividing the data and constructing a ratio that preserves statistical power whilst reducing sensitivity to slow rate changes~\cite{Run1PRDomegaa}.
Each fit models the data well, producing reduced $\chi^{2}$ values consistent with unity.
Fourier transforms of the fit residuals have no unexpected frequencies as shown in Fig.~\ref{fig:precession}.
Scans of fit start and end times, positron energy, and individual calorimeter stations show variation in \wam consistent with statistical expectations.
After unblinding, the analysis groups determine consistent values for \wam and their independently estimated systematic uncertainties.
We combine the six asymmetry-weighted methods equally for the final central value and verify the result with other less sensitive methods.

The extraction of \wam is the only aspect of the result with significant statistical uncertainty. The number of positrons above 1000~MeV entering the asymmetry-weighted analyses increased from $15 \times 10^9$ in \RunOne to $71 \times 10^9$ in \RunTwoThree. This reduces the statistical uncertainty from 434 ppb to 201 ppb.

The systematic uncertainty on \wam is also reduced by a factor greater than 2 to 25~ppb.
The largest reduction comes from our treatment of pileup, when two positrons enter a calorimeter close in time and are not separated by reconstruction algorithms.
The difference in phase between two lower-energy positrons and a single higher-energy positron, coupled with a rate change over the storage period, can bias \wam.
Each calorimeter comprises a $9 \times 6$ array of PbF$_2$ crystals that are read out independently.
Improved clustering of crystal hits in the reconstruction algorithms reduces the number of unresolved pileup events.
In addition, some groups adopted a method of overlaying waveforms rather than modeling the reconstruction response to proximate crystal hits.
The pileup uncertainty is reduced from 35~ppb in \RunOne to 7~ppb in \RunTwoThree.

The other significant reduction is related to transverse beam oscillations.
The repair of the damaged ESQ resistors removes the majority of systematic effects associated with large changes in the betatron frequencies over a muon storage period.
Additionally, the higher statistical precision allows for improved empirical modeling of the decoherence envelope, enabling a wider range of possibilities to be studied.
The uncertainty drops from 38~ppb in \RunOne to 21~ppb but remains the dominant systematic uncertainty for \RunTwoThree for \wam.

Smaller reductions are achieved in the systematic uncertainties from a residual early-to-late effect and the calorimeter gain correction (see Ref.~\cite{Run1PRDomegaa}), resulting in values of 10~ppb and 5~ppb, respectively.


\textit{Beam-dynamics corrections ${C_i}$}.---Five corrections must be made to convert the measured frequency \wam into the anomalous precession frequency \wa in Eq.~\ref{eq:amueq}.

The largest correction is due to the electric fields of the ESQs. The effect on \wa is minimized by the choice of nominal muon momentum \SI{3.1}{GeV/c}~\cite{e821quadpaper}. The electric field correction $C_e$ is required to account for the momentum spread of the muon beam. 

The muon momentum distribution is determined from the frequency distribution and debunching rate of the injected beam using calorimeter data. Additionally, the radial distribution of stored muons over a betatron period is obtained from tracker data. The debunching analysis takes into account differences in momentum spread along the injected bunch length that were not included in the \RunOne analysis.
Accounting for this difference and using complementary tracker information reduces the $C_e$ uncertainty from 52~ppb in \RunOne to 32~ppb in \RunTwoThree.

A pitch correction $C_p$ accounts for the reduction of \wa caused by vertical betatron oscillations.
We use tracker data to extract the distribution of vertical betatron amplitudes.
The analysis is largely unchanged from \RunOne.

Any temporal change to the muon ensemble-average phase $\varphi_0$ in Eq. 3 will bias \wam.  
Correlations between the muon decay position and $\varphi_0$ are accounted for through the phase acceptance correction $C_{pa}$.
This correction is evaluated by measuring the transverse beam distribution throughout the storage period and using simulations to determine the shifts in average phase at the calorimeters.
The size of $C_{pa}$ is determined by variation in the beam spatial distribution, which is significantly reduced by replacing the damaged ESQ resistors, and the associated systematic uncertainty is reduced from 75~ppb to 13~ppb.

Phase is also correlated with muon momentum owing to the momentum-dependent phase advance in upstream beamline components~\cite{Run1PRABBeamDyn}.
A differential decay correction $C_{dd}$ is required since the higher-momentum muons have a longer boosted lifetime than lower-momentum muons.
Three separate contributions to the $C_{dd}$ correction yield a $-15$~ppb correction with 17~ppb uncertainty.
This correction was not applied to the \RunOne analysis.

Muons lost during a storage period can also lead to a change in the muon momentum distribution.  This effect has also been greatly reduced by replacing the ESQ resistors. The correction factor $C_{ml}$ is evaluated as $0\pm3$~ppb compared to a correction in \RunOne of $-11\pm 5$~ppb. 

\textit{Muon-weighted magnetic field \weightedfield}.---The increased temperature stability in \RunTwo and \RunThree due to thermal magnet insulation and improved hall temperature stability results in a significantly more stable magnetic field (RMS of  \SI{2}{ppm} for \RunTwo and \SI{0.5}{ppm} for \RunThree).
Additional systematic measurements of the temperature dependence of the petroleum-jelly-based NMR probes 
used in the trolley have reduced the systematic uncertainty from trolley temperature changes to 9--15~ppb, depending on the data set.

The calibration procedure is improved for \RunTwoThree compared to \RunOne. Not only are two calibrations performed, one for each run, but the process is also optimized, resulting in reduced uncertainties.
Small differences between the sample volume in the calibration and the trolley probes are now corrected.
In addition, correction terms for the calibration probe are determined more precisely.
The overall systematic uncertainty from calibration is below 20~ppb.

As in \RunOne, the magnetic field is parametrized in a multipole expansion in transverse planes.
In the current analysis, the number of terms used has increases from $9$ to $12$, improving the fit quality.
The dominant uncertainties for the spatial field maps---each approximately \SI{20}{ppb} in magnitude---arise from NMR frequency extraction~\cite{RanFreqPaper}, the motion effects of the trolley, and the estimated perturbation by the mechanism used to retract the trolley from the storage region.

The systematic uncertainty of tracking the field in time using the fixed probe data between two field maps is estimated by a Brownian bridge model tuned to the observed mismatch from propagating one map to another.
Because of the larger number of field maps ($69$ in \RunTwoThree, compared to $14$ in \RunOne), the uncertainty from the field tracking is reduced to 10--16~ppb depending on the data set. We discovered and corrected a tracking bias as a function of time after the last magnet ramp-up (3--10~ppb). 

\begin{figure}[ht]
\centering
\includegraphics[width=\columnwidth]{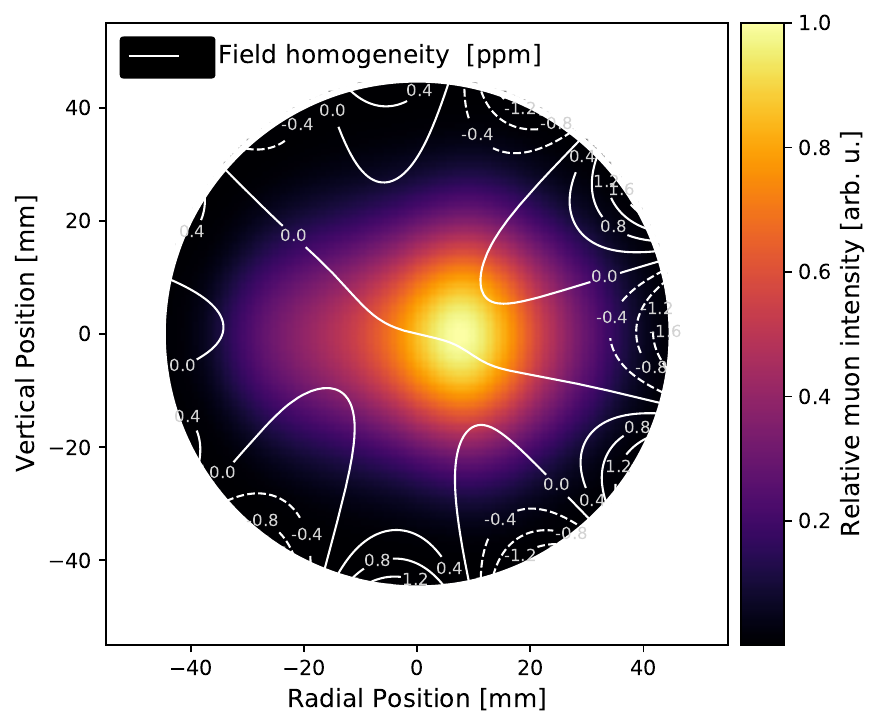}
\caption{Azimuthally averaged magnetic field contours overlaid on the time- and azimuthally averaged muon distribution for the \RunThreeB data set.  The field is more uniform, and the increased kicker strength moves the beam closer to the center than in \RunOne.} \label{fig:field}
\end{figure}

The muon weighting follows the same approach used in \RunOne. The more uniform field reduces the uncertainties by around a factor of 2 to 7--13~ppb. The beam distribution and azimuthally averaged magnetic field from \RunThreeB are shown in Fig.~\ref{fig:field}.

\textit{Magnetic field transients $B_i$}.---Transient magnetic fields synchronized with beam injection are caused by the pulsing of ESQs and eddy currents in the kickers.
Both effects require corrections to the muon-weighted magnetic field and are improved significantly compared to \RunOne by additional measurements.

In \RunOne, the correction from the magnetic field transient due to vibrations caused by ESQ pulsing, $B_q$, was only measured at a limited number of locations around the ring. Using the same vacuum-sealed petroleum-jelly-based NMR probe, but now on a nonconductive movable device, we mapped the transient fields in the storage region between the ESQ plates azimuthally.
This mapping, in combination with improved methodology and repeated measurements over time, leads to a reduction of the formerly dominant systematic effect by more than a factor of 4 to \SI{20}{ppb}.

The effect of kicker-induced eddy currents $B_k$ was measured with the same fiber magnetometer based on Faraday rotation in terbium gallium garnet crystals used in \RunOne~\cite{Run1PRAField}.
An improved setup, mainly to further reduce vibrations, and more extensive measurements, reduces the uncertainty by around a factor of $3$ to \SI{13}{ppb}.

\textit{Consistency checks}.---In addition to the three data subsets described here, the data are further subdivided based on a number of monitored experimental parameters to examine possible correlations. These parameters include ring temperature,  magnet current, vacuum pressure, day/night, time since magnet ramp-up and variables associated with the beam motion.
We find no statistically significant correlations between our results and any of these parameters. 

\textit{Calculation of $a_\mu$}.---Table \ref{tb:Rmu} contains the values of \wa and \opprimetilde, including all correction terms in Eq.~\ref{eq:Rcomponents}, for the three data subsets and their ratios \Rmuprime.
The statistical uncertainty dominates in each subset, and as such, the \Rmuprime values are largely uncorrelated.
Nearly all systematic uncertainties that enter into \Rmuprime are fully correlated across the subsets.
Over the course of this analysis, three small errors in the \RunOne analysis were identified \footnote{We have updated the \RunOne measurement with three corrections. The E-field correction $C_e$ was inadvertently calculated using the blinded clock frequency. We have updated this to use the unblinded clock frequency. This changes the \RunOne result by $+19$~ppb. The phase acceptance correction $C_{pa}$ for \RunOneB was inadvertently swapped with the $C_{pa}$ for \RunOneC. We have applied these correctly now. This changes the \RunOne result by $+6$~ppb. A correction was applied to the temperature dependence of the magnetic susceptibility of a spherical water probe. The applied correction did not include an additional term that accounts for the temperature dependence of the density of water. We have now included this term. This changes the \RunOne result by $+2$~ppb. These corrections are all positive and sum to a $+28$~ppb correction to the \RunOne result.}. The total shift in the previous result due to these errors is $+28$~ppb, which has been applied to the value reported in this Letter.

The weighted-average value of the \RunTwoThree data is \RmuprimeTwoThree = 0.00370730088(75)(26), where the first error is statistical and the second is systematic.
This value is in excellent agreement with the adjusted \RunOne value \RmuprimeOne = 0.0037073004(16)(6). 
Assuming that the systematic errors are fully correlated between \RmuprimeTwoThree and \RmuprimeOne, we obtain the combined value of \RmuprimeOneTwoThree = 0.00370730082(68)(31).

\begin{figure}[ht]
\centering
\includegraphics[width=\columnwidth]{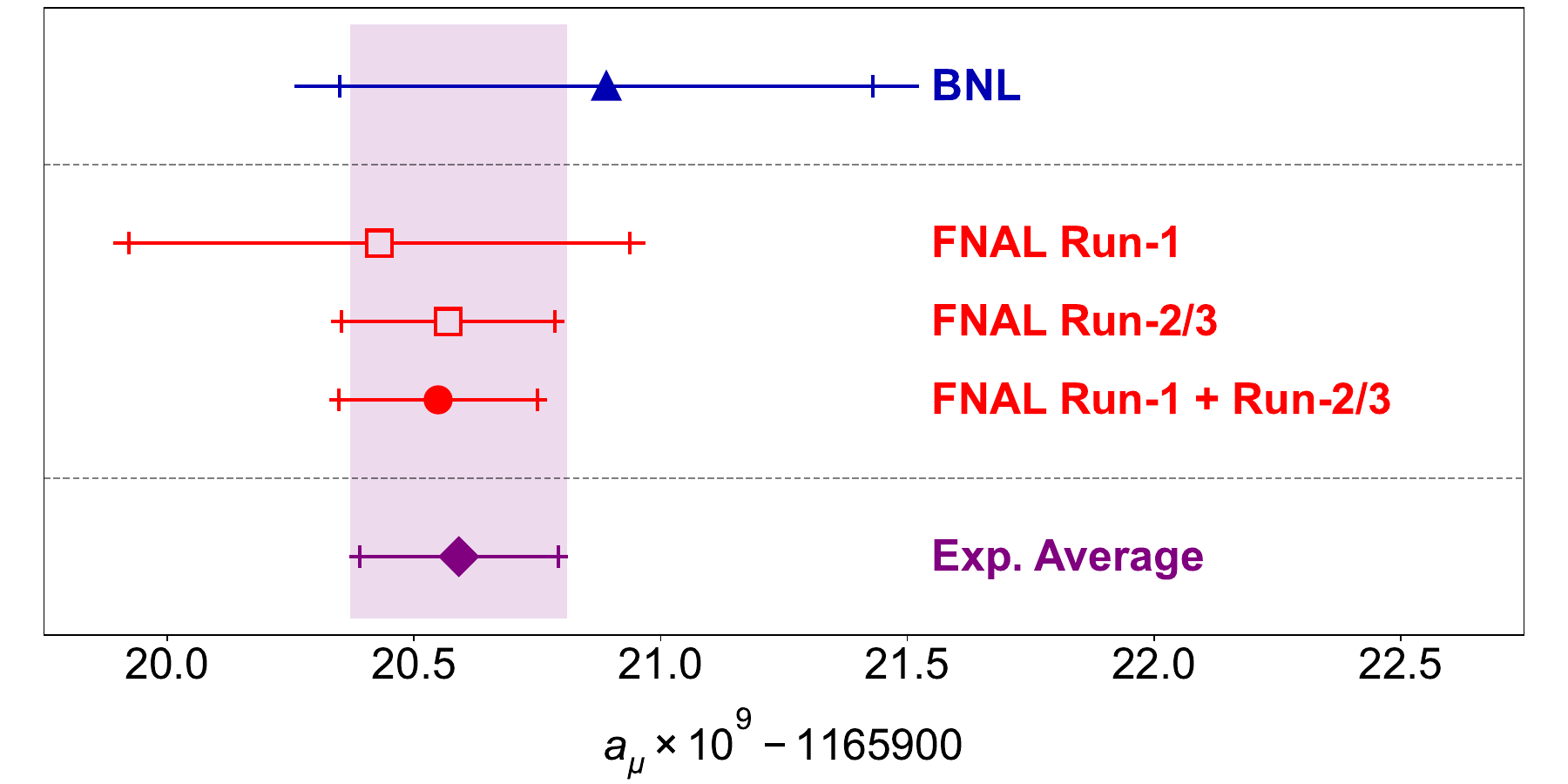}
\caption{Experimental values of \amu from BNL E821~\cite{BNLFinalReport}, our \RunOne result~\cite{Run1PRL}, this measurement, the combined Fermilab result, and the new experimental average. The inner tick marks indicate the statistical contribution to the total uncertainties.} \label{fig:results}
\end{figure}

From Eq.~\ref{eq:amueq}, we arrive at a new determination of the muon anomaly,
\begin{equation*}
a_\mu({\rm FNAL}) = 116\,592\,055(24) \times  10^{-11}  ~~~ (\text{0.20\,ppm}),
\end{equation*}
where the statistical, systematic, and external parameter uncertainties from Table~\ref{tb:systematics} are combined in quadrature.
The combined (BNL and FNAL) experimental (exp) average becomes
\begin{equation*}
\amu(\text{exp}) = 116\,592\,059(22) \times 10^{-11}   ~~~(0.19\,\text{ppm}).
\end{equation*}
The results are displayed in Fig.~\ref{fig:results}.

A comprehensive prediction for the SM value of the muon magnetic anomaly was compiled most recently by the Muon \gmtwo Theory Initiative in 2020~\cite{TI}, using results from Refs.~\cite{Aoyama:2012wk,Aoyama:2019ryr,Czarnecki:2002nt,Gnendiger:2013pva,Davier:2017zfy,Keshavarzi:2018mgv,Colangelo:2018mtw,Hoferichter:2019mqg,Davier:2019can,Keshavarzi:2019abf,Kurz:2014wya,Melnikov:2003xd,Masjuan:2017tvw,Colangelo:2017fiz,Hoferichter:2018kwz,Gerardin:2019vio,Bijnens:2019ghy,Colangelo:2019uex,Blum:2019ugy,Colangelo:2014qya}. The leading-order hadronic contribution, known as hadronic vacuum polarization (HVP), was taken from $e^+e^-\rightarrow$ hadrons cross-section measurements performed by multiple experiments. However, a recent lattice calculation of HVP by the BMW Collaboration~\cite{BMW} shows significant tension with the $e^+e^-$ data.  In addition, a new preliminary measurement of the $e^+e^-\rightarrow \pi^+\pi^-$ cross section from the CMD-3 experiment~\cite{CMD3} disagrees significantly with all other $e^+e^-$ data. There are ongoing efforts to clarify the current theoretical situation~\cite{TISnowmassPaper}. While a comparison between the Fermilab result from \RunOneTwoThree presented here, $a_\mu({\rm FNAL})$, and the 2020 prediction yields a discrepancy of $5.0 \sigma$, an updated prediction considering all available data will likely yield a smaller and less significant discrepancy.

\begin{table}
\begin{ruledtabular}
\begin{tabular}{lrrr}
Run & $\omega_a/2\pi$\,[Hz] & $\opprimetilde/2\pi$\,[Hz] & $\Rmuprime \times 1000$\\
\hline
\RunOne &  &  & 3.7073004(17)\\
\hline
\RunTwo & 229077.408(79) & 61790875.0(3.3) & 3.7073016(13)\\
\RunThreeA & 229077.591(68) & 61790957.5(3.3) & 3.7072996(11)\\
\RunThreeB & 229077.81(11) & 61790962.3(3.3) & 3.7073029(18)\\
\hline
\RunTwoThree&  & \multicolumn{2}{r}{3.70730088(79)}\\
\hline
\multicolumn{2}{l}{\RunOneTwoThree} & \multicolumn{2}{r}{3.70730082(75)}\\
\end{tabular}
\end{ruledtabular}
\caption{Measurements of \wa, \opprimetilde, and their ratios \Rmuprime multiplied by 1000. The \RunOne value has been updated from~\cite{Run1PRL} as described in the text.}
\label{tb:Rmu}
\end{table}

In summary, we report a measurement of the muon magnetic anomaly to $0.20$ ppm precision using our first three years of data.  This is the most precise determination of this quantity, and it improves on our previous result by more than a factor of 2.  Analysis of the remaining data from three additional years of data collection is underway and is expected to lead to another factor of 2 improvement in statistical precision.

We thank the Fermilab management and staff for their strong support of this experiment, as well as our university and national laboratory engineers, technicians, and workshops for their tremendous support.
Greg Bock and Joe Lykken set the blinding clock and diligently monitored its stability.

The Muon \gmtwo Experiment was performed at the Fermi National
Accelerator Laboratory, a U.S. Department of Energy, Office of
Science, HEP User Facility. Fermilab is managed by Fermi Research
Alliance, LLC (FRA), acting under Contract No. DE-AC02-07CH11359.
Additional support for the experiment was provided by the Department
of Energy offices of HEP and NP (USA), the National Science Foundation
(USA), the Istituto Nazionale di Fisica Nucleare (Italy), the Science
and Technology Facilities Council (UK), the Royal Society (UK),
the National Natural Science Foundation of China
(Grants No. 11975153 and No. 12075151), MSIP, NRF, and IBS-R017-D1 (Republic of Korea),
the German Research Foundation (DFG) through the Cluster of
Excellence PRISMA+ (EXC 2118/1, Project ID 39083149),
the European Union Horizon 2020 research and innovation programme under
the Marie Sk\l{}odowska-Curie Grant Agreements No. 101006726 and No. 734303,
the European Union STRONG 2020 project under Grant Agreement No. 824093,
and the Leverhulme Trust, LIP-2021-01.

\authornote{Deceased}

\end{document}